\documentclass[10pt, compsocconf]{IEEEtran}
\usepackage{times}
\usepackage{algorithm}
\usepackage{algorithmic}
\usepackage{adjustbox}
\usepackage{todonotes}
\usepackage{amsmath}
\usepackage{amsfonts}
\usepackage{xcolor}
\usepackage{balance}

\usepackage{graphicx}
\usepackage{subfig}

\usepackage{xcolor}

\usepackage{listings}
\usepackage {parcolumns}

\definecolor{verylightgray}{gray}{0.85}

\usepackage{balance}
\makeatletter
\def\@copyrightspace{\relax}
\makeatother

\title{User-transparent Distributed TensorFlow}

\author{
Abhinav Vishnu{\small $~^{\#1}$},
Joseph Manzano{\small $~^{\#3}$},
Charles Siegel{\small $~^{\#2}$}, and
Jeff Daily{\small $~^{\#4}$}
\vspace{1.6mm}\\
\fontsize{10}{10}\selectfont\itshape
$^{\#1,2,3,4}$\,Pacific Northwest National Laboratory,
Richland, WA 99352\\
}

\begin{document}
\maketitle

\begin{abstract}
Deep Learning (DL) algorithms have become the {\em de facto} choice for data
analysis. Several DL implementations -- primarily limited to a single compute
node -- such as Caffe, TensorFlow, Theano and Torch have become readily
available. Distributed DL implementations capable of execution on large scale
systems are becoming important to address the computational needs of large data
produced by scientific simulations and experiments.  Yet, the adoption of
distributed DL implementations faces significant impediments: 1) most
implementations require DL analysts to modify their code significantly -- which
is a show-stopper, 2) several distributed DL implementations are geared towards
cloud computing systems -- which is inadequate for execution on massively
parallel systems such as supercomputers. 

This work addresses each of these problems. We provide a distributed memory DL
implementation by incorporating required changes in the TensorFlow runtime
itself.  This dramatically reduces the entry barrier for using a distributed
TensorFlow implementation.  We use Message Passing Interface (MPI) -- which
provides performance portability, especially since MPI specific changes are
abstracted from users. Lastly -- and arguably most importantly -- we make our
implementation available for broader use, under the umbrella of Machine
Learning Toolkit for Extreme Scale (MaTEx) at {\texttt http://hpc.pnl.gov/matex}. We
refer to our implementation as MaTEx-TensorFlow.

\end{abstract}

\section{Introduction}
Machine Learning and Data Mining (MLDM) algorithms are becoming quintessential
in analyzing large volume of data produced by simulations, experiments and
mobile devices~\cite{data:ascac11,data:ascac13}.  MLDM algorithms are generally
divided into {\em supervised} (the input data set is labeled with the ground
truth) and {\em unsupervised} (learning from unlabeled data) algorithms.  Base
supervised/unsupervised algorithms may be combined together using {\em
ensemble} methods.  Several software packages that support supervised,
unsupervised and ensemble algorithms have become available including
Weka~\cite{weka}, Scikit~\cite{scikit}, libsvm~\cite{libsvm}, and
Matlab~\cite{MATLAB:2010}. 

Deep Learning (DL) algorithms are a class of MLDM algorithms that emulate
the computational structure of a mammalian brain  by using several layers of {\em
neurons} interconnected with {\em synapses} and learn the weights for the
synapses using gradient descent method.  DL algorithms can be divided into
several classes: Multi-Layer Perceptrons (MLP - typically used on tabular data
sets), Convolutional Neural Networks (CNNs - typically used on images and other
spatially related data) and Recurrent Neural Networks (RNNs - typically used on
sequential and time-series data).  Many researchers have applied DL algorithms
to solve problems in their domains, often reporting better results than the
state of the art published models.  These domains include high energy
physics~\cite{Baldi:2014kfa}, computational biology~\cite{ben2008support} and
cyber security~\cite{tarca:bio07,vossen:hep08,ml:climate, liu2016application}.
Naturally, open source toolkits such as Theano~\cite{bergstra+al:2010-scipy,
Bastien-Theano-2012}, Torch~\cite{Collobert02torch:a} and Caffe~\cite{caffe} which use cuDNN~\cite{chetlur2014cudnn} have become widely available.
\begin{figure}[hptb]
	\centering
	\includegraphics[width=0.9\columnwidth]{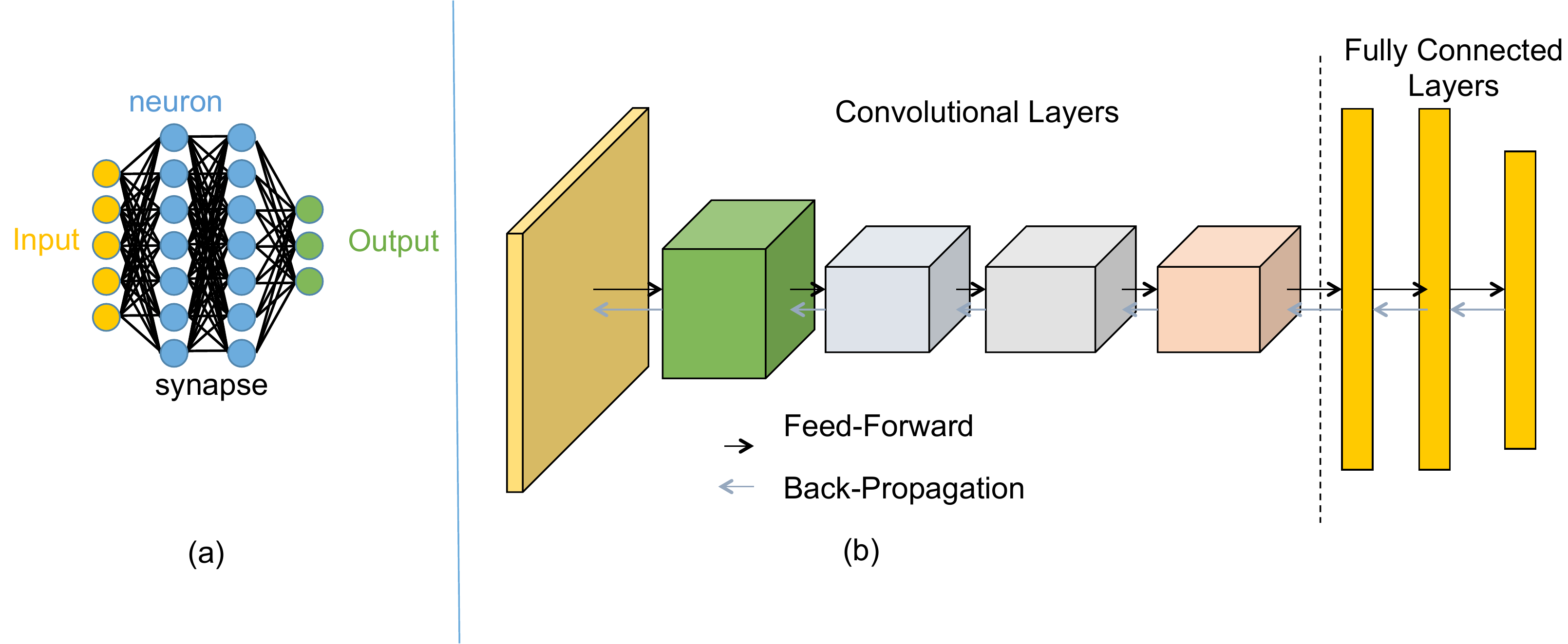} 
	\caption{(a) A pictorial representation of a neural network with two hidden layers (b) An example of a neural network -- AlexNet~\cite{NIPS2012_4824}. AlexNet has two types of layers: 1) convolutional layers for extracting features from images, 2) fully connected layers for using these features for classification}
	\label{fig:dnn}
\end{figure}

In November 2015, Google released TensorFlow, an open source toolkit for
developing MLDM algorithms primarily suited to implementing DL algorithms.  It
uses a dataflow model by specifying operations on tensors (multi-dimensional
arrays).  TensorFlow supports automatic differentiation, which simplifies the
design and implementation of gradient descent methods for novel structures.
This allows TensorFlow to readily support MLPs, CNNs and RNNs on
multi-core/many-core systems (GPUs) and supports the use of algorithmic
improvements, such as AdaGrad~\cite{Duchi}, Adam~\cite{kingma2014adam} and
Momentum~\cite{sutskever2013importance} gradient descent and neuron dropout for
regularization~\cite{hinton2012improving}.  

Distributed TensorFlow (starting with version 0.8.0) has become available for
execution on multiple nodes. These compute nodes may possibly be connected with
multiple GPUs on each node. Google's distributed TensorFlow is based on
Google's RPC (gRPC), which is primarily geared towards cloud computing systems
interconnected using Ethernet. This is inadequate for supercomputers, which
typically use interconnects such as InfiniBand, Intel Omni-path and Cray
interconnects for leveraging high bandwidth and Remote Direct Memory Access
(RDMA) features. A few efforts such as gRPC using Message Passing Interface
(MPI)~\cite{mpi1, mpi2} have attempted to address this limitation.
Besides limited applicability to HPC interconnects, gRPC is primarily geared towards parameter sever based DL
implementations, which diverges from the convergence properties of sequential
batch/stochastic gradient descent (SGD). Recently, Baidu announced the
availability of MPI with TensorFlow by introducing a novel All-to-all reduction
technique and user-operations which may be added to existing TensorFlow
scripts. While optimized for performance, Baidu's contributions require several
changes related to MPI in existing TensorFlow scripts.

At the same time, the majority of DL analysts tend to write a sequential
TensorFlow program. This leads to our problem statement: {\em Can we design a
TensorFlow runtime capable of execution on multiple nodes without requiring any
TensorFlow specific changes to existing scripts?} 

\subsection{Contributions}
Specifically, we make the following contributions in this paper: 

\begin{itemize}
	\item We consider several design choices for implementing distributed
		TensorFlow such as defining new user-operations, and methods to
		synchronize replicas (since we focus on data parallelism) 	
	\item We evaluate our implementation on two platforms: 1) Intel
		multi-core system connected with InfiniBand, and 2) NVIDIA
		multi-GPU system connected with InfiniBand

	\item We provide our implementation by extending TensorFlow 1.0 for
		broader use by making it available under the umbrella of
		Machine Learning Toolkit for Extreme Scale
		(MaTEx)~\cite{1603.02339,matex}. We refer to our implementation as MaTEx-TensorFlow.
\end{itemize}

We observe that MaTEx-TensorFlow scales well on multiple compute nodes using
ImageNet LSVRC12 datasets and AlexNet, GoogLeNet, InceptionV3 and ResNet-50 neural
network topologies. Our primary contribution is the ability to leverage the
multi-node CPU systems, and multi-node GPU implementations, without modifying
any source code specific to TensorFlow. We recommend using our data readers,
which provide a simple interface for reading data available in multiple formats.

The rest of the paper is organized as follows: In section~\ref{sec:background},
we present the background of our work. In section~\ref{sec:design}, we present a
solution space for designing MaTEx-TensorFlow.  We present an in-depth performance
evaluation in section~\ref{sec:exp}, followed by related work in
section~\ref{sec:related} and conclusions in section~\ref{sec:conclusions}.

\section{Background}
\label{sec:background}
In this section, we provide a brief background of Google
TensorFlow~\cite{tensorflow2015-whitepaper} (simply referred as TensorFlow for
rest of the paper) and Message Passing Interface (MPI)~\cite{mpi1, mpi2}.

\subsection{TensorFlow} 
Google released TensorFlow in November 2015 as a platform for building and
developing DL implementations. TensorFlow is capable of utilizing multiple threads,
such that multi-core systems can be utilized effectively. It also provides
implementations to leverage GPUs (using NVIDIA CUDA based DNN (cuDNN)), such that one (or more) GPUs
on a single node may  be utilized effectively.

\subsubsection{TensorFlow Graph}
The fundamental model of computation within TensorFlow is a {\em computational
graph}.  A graph contains vertices, representing {\em operations}, and edges,
representing {\em tensors} (arbitrary dimensional arrays).  Each operation can take
multiple inputs and generate multiple outputs, with tensors created and passed
from one operation to another.  Edges also act as control flow objects
in the computational graph, which ensures dependencies, that naturally arise in
DL implementations.

\subsubsection{Tensors}
There are several special types of tensors in TensorFlow.  An important tensor is a
{\em variable}.  Variables are persistent tensors that can be accessed outside
the computational graph. In DL implementations, the {\em weights} and {\em
biases} of a model are stored as variables and updated by operations, when a
computational graph is executed.  
Another type of a tensor is {\em placeholder}.  Placeholders are input points
into a computational graph.  Outside of placeholders, the computational graph
is self-contained.  

\subsubsection{Session}
In TensorFlow, a {\em session} controls the graph.  It stores the values of variables and is used to run the computations described by the graph.  After the creation of a session, an {\em initializer} must be run to give values to the variables to be used within the session.  Subsequent computations, such as the computation of gradients, must be managed through the session to ensure that the correct values of variables are used.  The session makes use of a {\em scheduler}, which maintains a record of which operations have been completed and enqueues those whose dependencies are all satisfied to be executed.

\subsubsection{Device Scheduling}
In addition to its use by the session to keep track of which operations are ready to execute, the TensorFlow scheduler also handles device scheduling when multiple devices are available.  Before executing a graph as desired by the user, the schedule runs a simulation of the graph to determine execution time and the order of the operations.  It then uses this information to create the dependency lists that the session requires and to assign each operation to a device.  These assignments first depend on whether there is an implementation of the operation for a given device -- for instance, sometimes GPU implementations may be unavailable -- and then upon expected execution speed taking into account inter-device communication times for the relevant tensors.

\subsection{Message Passing Interface} Message Passing Interface
(MPI)~\cite{mpi1,mpi2} provides a rich set of abstractions for inter-process
communication. It supports pair-wise communication (such as using send,
receive) and group communication (such as using reduction, barrier). MPI has
become the {\em de facto} communication interface for legacy scientific
applications.
The primary reason for MPI's success is its wide availability. MPI is available
on large scale supercomputers, cloud computing systems and it can also be used
for inter-process communication on a single compute node -- if other shared
memory programming models are not available.  Unlike other runtimes such as
Spark and gRPC, MPI is able to take advantage of high performance interconnects
such as InfiniBand, Intel Omni-Path and Cray interconnects interconnects effectively.
Due to the performance reasons, we considered MPI to be the primary
communication interface instead of other communication subsystems. 

In MaTEx-TensorFlow, we have used several MPI routines for our large scale
implementation. We have used All-to-all reduction (an MPI primitive which
allows operations such as sum on user's data, and disseminates the final result
among all the processes in a group) for averaging gradients and point-to-point
operations for data distribution.
We also observed that MPI has been criticized for its lack of support for fault
tolerance. However, with recent advancements -- such as User-level Fault
Mitigation (ULFM) -- and open source implementations, it is possible to design
fault tolerant DL algorithms using MPI, without losing performance and
"continued execution" in the presence of hardware faults. We expect that with
ULFM (or its variants) becoming available with mainstream implementations, MPI
would find its wide acceptance in the DL community.

\section{MaTEx-TensorFlow Design Space}
\label{sec:design}
In this section, we present a detailed description of MaTEx-TensorFlow design space.

\begin{figure*}[!htbp]
\subfloat{\includegraphics[width=\textwidth]{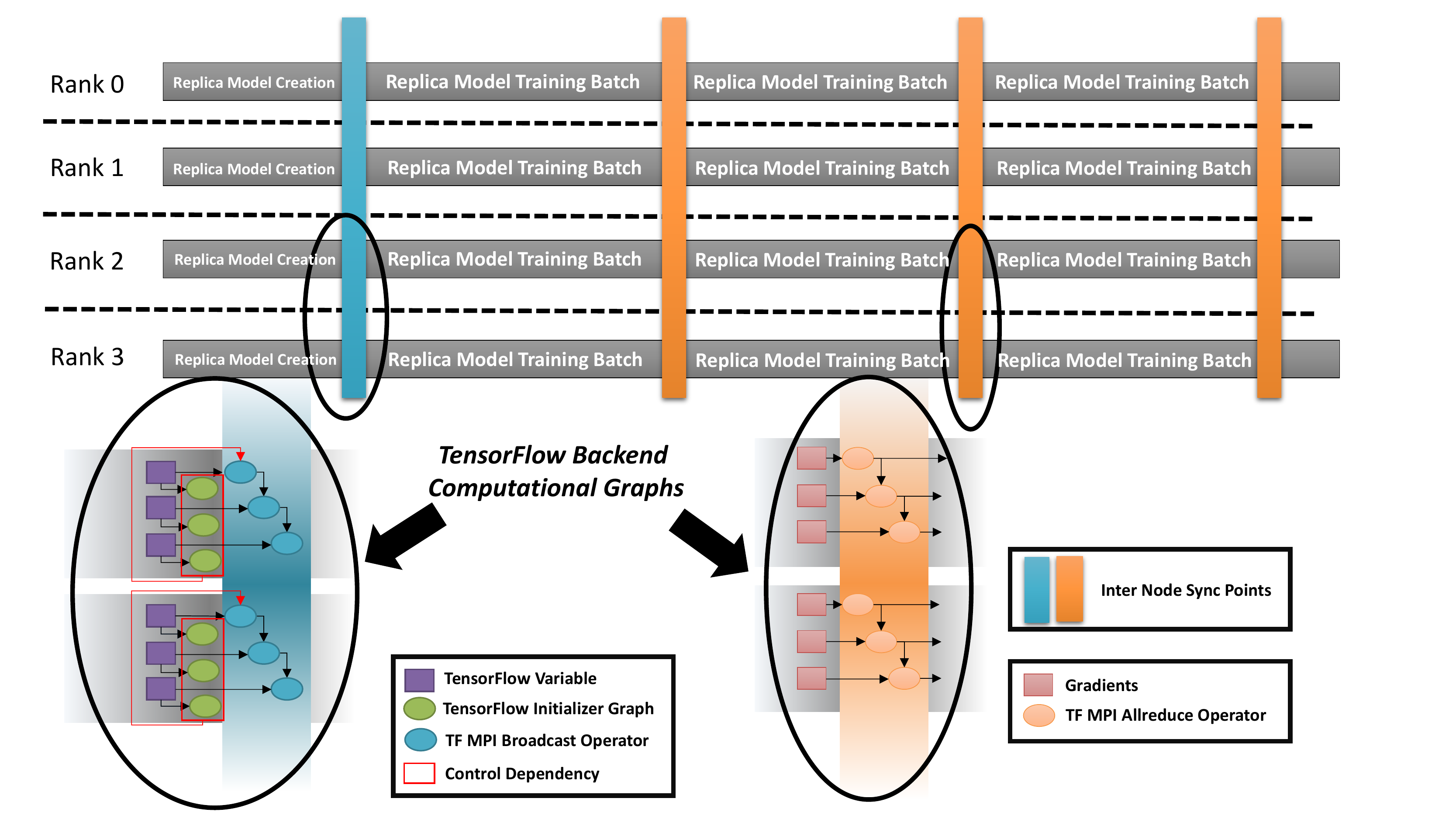}}
\caption{Example of a MaTEx TensorFlow executing on four MPI ranks. Each rank will run a model replica and communicate at each of the reduction points (i.e. the orange bars). Each model is initialized identically due to the broadcast operator at the beginning (i.e the blue bar).}
\label{fig:figml}
\end{figure*}

\subsection{Data Parallelism/Model Parallelism}
An important design consideration is the type of parallelism to be used for
MaTEx-TensorFlow. In model parallelism, the layers in a DNN are split across
multiple devices (such as GPUs and/or multiple compute nodes). The model
parallelism is potentially effective in scale-out, since the scheduling on
multiple devices enables the use of small batch sizes. 

However,  DNNs increasingly contain deeper convolutional layers, where the size
of the activations is much larger than the overall model. Under model
parallelism, these activations would need to be communicated across devices --
which is prohibitive.  Hence, it is worthwhile to consider {\em data
parallelism}, where the model is replicated and the data is split across multiple compute devices.
Similar observations have been pointed out by Krizhevsky {\em et
al.}~\cite{oneweirdtrick}. Hence, we use data parallelism for implementing
MaTEx-TensorFlow. 

\subsection{Programming Models}
We considered several programming models/interfaces for implementing
MaTEx-TensorFlow. Specifically, we considered Spark, Hadoop, gRPC and MPI.
MapReduce frameworks such as Spark~\cite{spark} and Hadoop~\cite{hadoop}
abstract the details of parallelism effectively. However, they are not suitable
for large scale systems which are typically connected using high performance
interconnects. 

Another possibility is to use Google's Remote Procedure Call (gRPC). The
initial implementation uses sockets interface, which is not suitable for HPC
interconnects. Recent implementations of gRPC using Remote Direct Memory Access
(RDMA) alleviate this limitation. However, the primary gRPC primitives do not
include all-to-all reduction based collective operations -- which is
problematic for scaling out SGD. gRPC is specifically targeted for
parameter-server (PS) based implementation of SGD. However, PS based
implementations suffer from slow convergence and communication bottlenecks.

An alternative choice is to use Message Passing Interface (MPI). It provides a
rich set of communication primitives including point-to-point, collective and
other operations. MPI is also widely available on large scale systems including
supercomputers, and cloud computing systems. For these reasons, we use MPI as
the communication interface for implementing MaTEx-TensorFlow.  MPI has
frequently been criticized due to lack of fault tolerance. While
MaTEx-TensorFlow is not fault tolerant, we plan to  handle fault tolerance for
MPI using ULFM -- which allows the MPI application to continue executing in the
presence of faults.  By using data parallelism  the critical data structures
are automatically replicated for fault tolerance.  This approach would allow
MPI to address the limitations of Spark while maintaining many of its
advantages. However, fault tolerant TensorFlow is beyond the scope of this paper.

\subsection{Existing Approaches for Distributed Memory}
Up to now, we have identified using MPI for implementing distributed memory
DL and data parallelism for scaling out the algorithms. It is equally important
to consider the level of abstraction which should be provided to the user.
There are several design choices 

\subsubsection{MPI-enabled TensorFlow Scripts} One possibility is to use MPI
within TensorFlow scripts -- visible to the end-user. This approach requires no
changes to the TensorFlow runtime, which makes it an attractive choice.  In the
previous version of MaTEx-TensorFlow, this approach was used~\cite{1603.02339}.
The upside of this approach is that a user who does not want to write
TensorFlow code may use these scripts to build DNNs.  However, in many cases,
users tend to write their customized TensorFlow scripts. Hence, they would be
required to add MPI specific changes in their code -- which is problematic for
these users.

\subsubsection{Class Packages} Another possibility is to create a module of helper functions and classes.  These functions and classes may then be used by TensorFlow users.  Recently, Baidu~\cite{baiduflow} has proposed work on this model.  Baidu's extensions are integrated into TensorFlow.  However, the user must still make Baidu-specific changes to their TensorFlow scripts to make use of these extensions for distributed memory execution.

\subsection{Proposed Approach for Distributed Memory}
We have observed that -- due to pre-existing, complex scripts -- the distributed 
memory implementations are inadequate for most DL analysts.
Hence, it is important to
consider implementations which would provide distributed memory DL while abstracting the changes from the users completely. That
is the focus of MaTEx-TensorFlow. In this section, we provide implementation
details along these lines.

For achieving this objective, we leverage TensorFlow operators. These operators
can be user-defined and inserted in the computational graph.  As shown in
Figure~\ref{fig:figml}, MaTEx-TensorFlow provides two new TensorFlow operators:
a \emph{Global Broadcast} for TensorFlow model variables and an
\emph{MPI\_Allreduce} operator for the model results (gradients) for the training
phase.  Both operators enhance the TensorFlow framework to provide support for
synchronous, data parallel models on a distributed memory system. 

\subsubsection{Broadcast Operator} 
MaTEx-TensorFlow ensures that each model replica is exactly the same at the
start of the training phase. To ensure this, we use a broadcast operator in
which the default MPI process (also referred as {\em rank} zero in MPI
terminology broadcasts the model at the start of the training phase.  A
TensorFlow variable has two components: 1) a tensor with actual value, and 2)
an associated computational graph operation.  For the broadcast operator, TensorFlow
creates an unordered list of initializer graphs for each variable. 
Since TensorFlow scheduler is unordered in scheduling variables, we add
explicit data dependencies to ensure that the buffers for broadcast are matched
correctly.

\subsubsection{MPI\_Allreduce Operator}
MaTEx-TensorFlow provides equivalence to the default SGD algorithm. Since it
uses data parallelism, the replicas need to be synchronized after each batch.
We use an MPI\_Allreduce operator for achieving this objective.  Since the gradients
(model updates) are returned as data tokens to the framework, the
MPI\_Allreduce operator has a simpler structure. 
The current version of MaTEx-TensorFlow provides layer-wise all-to-all
reduction.  This sets up an ordered list of reduction operators and then sequentially
synchronizes each layer across ranks, ensuring that the buffers are correctly ordered.

The use of {MPI\_Allreduce} function provides a communication complexity
$O(\log(p))$, where $p$ is the number of nodes.  As the work to compute the
gradients is divided evenly among nodes when using strong scaling, this will
provide approximately $C/p + log(p)$ work, where $C$ is the amount of
computation necessary to compute the gradients for each batch on a single
compute node.

\subsubsection{User-operations versus TensorFlow Runtime}
We choose to modify the TensorFlow backend directly.  Though this has an
increased engineering requirement, it allows for delivering a seamless user
experience.  Very few changes are required for the user's scripts in this
schema, making this method the simplest for the end-user, with the only
substantial changes being the use of parallel data readers rather than
sequential ones.

\subsection{Synchronous versus Asynchronous Implementation}

To enable efficient implementation of the backend modifications, we place
certain constraints on how data is distributed across the system.  The most
significant constraints are that data parallelism is the only mode that will be
used and that synchronous algorithms are the main vehicles of computation.

The choice of data over model parallelism is due to the trend towards more
expensive computation and fewer parameters for state-of-the-art neural
networks.  Model parallelism distributes different pieces of the model across
different nodes, and for a DL algorithm transmits the activations, which are
large for convolutions and small for fully connected networks.  Data
parallelism, however, duplicates the model across nodes and divides up the
processing of the dataset between them.  For convolutions, this is far more
efficient~\cite{krizhevsky2014one}.  Moreover, as we are requiring that our
algorithms be synchronous, the advantages of model parallelism decrease
further.

We implement synchronous models rather than asynchronous models to maintain
numerical equivalence with the sequential algorithm (c.f.
Figure~\ref{fig:losses}).  Synchronous models maintain this equivalence, but at
the cost of potentially having some devices idle at times.  Asynchronous models
prioritize full utilization of all devices at all times over equivalence to the
sequential algorithm.  A way in which asynchronous algorithms are used is under
the parameter server paradigm, where a single node is responsible for
maintaining the model and the remaining nodes are workers.  Each worker
independently computes updates which are applied by the model as they are
received.  This paradigm might leads to stale updates, and in many cases
requires a ``warm start,'' that is, for the model to be trained synchronously
for a time before switching to a parameter server.  At large scale, the
server/worker model can create a communication bottleneck as well where the
server(s) are overwhelmed with worker requests.

\subsection{I/O Considerations and Data Readers}
Besides supporting user-transparent distributed memory execution, MaTEx
provides interfaces for reading and automatically distributing datasets across
multiple compute nodes. Currently, MaTEx supports parallel NetCDF format, CSV,
MNIST and CIFAR dataset formats.

\subsection{Putting It All Together}
In this section, we present the integration of the proposed MaTEx-TensorFlow
design. Specifically, we have extended TensorFlow 1.0.0 for this purpose. The
changes regarding the runtime are completely abstracted from the user. 
As shown in Figure \ref{fig:codech}, the difference between the serial TensorFlow
script and multi-node script are only related to data readers. These readers are
considered optional as well. The only requirement is to provide input numpy
arrays.  

\label{sec:codechanges}
\begin{figure*}[!htbp]
\begin{parcolumns}{2}
\colchunk[1]{
\begin{lstlisting}
import tensorflow as tf 
import numpy as np
...
<@\textcolor{red}{from datasets import DataSet}\label{line_import}@>
... 
<@\textcolor{red}{image\_net = DataSet(\ldots)}\label{load_distributed}@>
data = <@\textcolor{red}{image\_net.training\_data}\label{alias_import1}@>
labels = <@\textcolor{red}{image\_net.training\_labels}\label{alias_import2}@>
...
# Setting up the network
...
# Setting up optimizer
...
init = tf.global_variables_initializer() 
sess = tf.Session() 
sess.run(init) 
...
# Run training regime
\end{lstlisting}
}\colchunk[2]{
\begin{lstlisting}
import tensorflow as tf 
import numpy as np
... 

...

data = ...   # Load training data
labels = ... # Load Labels
...
# Setting up the network
...
# Setting up optimizer
...
init = tf.global_variables_initializer()
sess = tf.Session()
sess.run(init)
...
# Run training regime
\end{lstlisting}}
\end{parcolumns}
\caption{(Left) A sample MaTEx-TensorFlow script, (Right) Original TensorFlow script. Notice that MaTEx-TensorFlow requires no TensorFlow specific changes.}
\label{fig:codech}
\end{figure*}

\section{Experimental Evaluation}
\label{sec:exp}
In this section, we present a performance evaluation of MaTEx-TensorFlow. We
compare the performance with serial TensorFlow. Table~\ref{table:arch} provides
a description of the architectures used for evaluation.
Table~\ref{table:datasets} provides a description of the datasets and neural
networks used for performance evaluation.

\begin{table*}[!htbp] 
		\centering
		\begin{tabular}{|c|c|c|c|c|c|c|c|c|}
				\hline
				Name  & CPU (\#cores) & GPU & Network & MPI & cuDNN & CUDA & Nodes & \#cores\\
				\hline 
				\bf{K40} & Haswell (20) & K40  & IB & OpenMPI 1.8.3 & 4 &  7.5 & 8 & 160\\
				\bf{SP} & Ivybridge (20) & N/A  & IB & OpenMPI 1.8.4 & N/A &  N/A & 20 & 400\\
				\hline
		\end{tabular}\\  
		\caption{Hardware and Software Description. IB (InfiniBand). The
		proposed research extends Baseline-Caffe incorporating
		architecture specific optimizations provided by vendors.}
		\label{table:arch}  
\end{table*}

\begin{table*}[!t] 
	\centering
	\begin{adjustbox}{max width=\textwidth}
		\begin{tabular}{|c|c|c|c|c|c|c|c|}
			\hline
			Dataset  & Neural Network     & Description & Training Samples & Validation Samples & Image Size            & Classes  \\
			\hline 
			ImageNet~\cite{ILSVRC15} & AlexNet~\cite{NIPS2012_4824} & Diverse Images & 1281167          & 50000              & $256\times256\times3$  & 1000  \\ 
			ImageNet & GoogLeNet~\cite{43022} & Diverse Images & 1281167          & 50000              & $256\times256\times3$ & 1000   \\
			ImageNet & InceptionV3~\cite{szegedy2016rethinking} & Diverse Images & 1281167          & 50000              & $256\times256\times3$ & 1000   \\
			ImageNet & ResNet50~\cite{he2016deep} & Diverse Images & 1281167          & 50000              & $256\times256\times3$ & 1000   \\
			\hline
		\end{tabular}
	\end{adjustbox}
	\caption{Datasets and neural networks used for performance evaluation}
	\label{table:datasets}  
\end{table*}

\subsection{Preliminaries}
\begin{figure}[!htbp]
\includegraphics[width=\columnwidth]{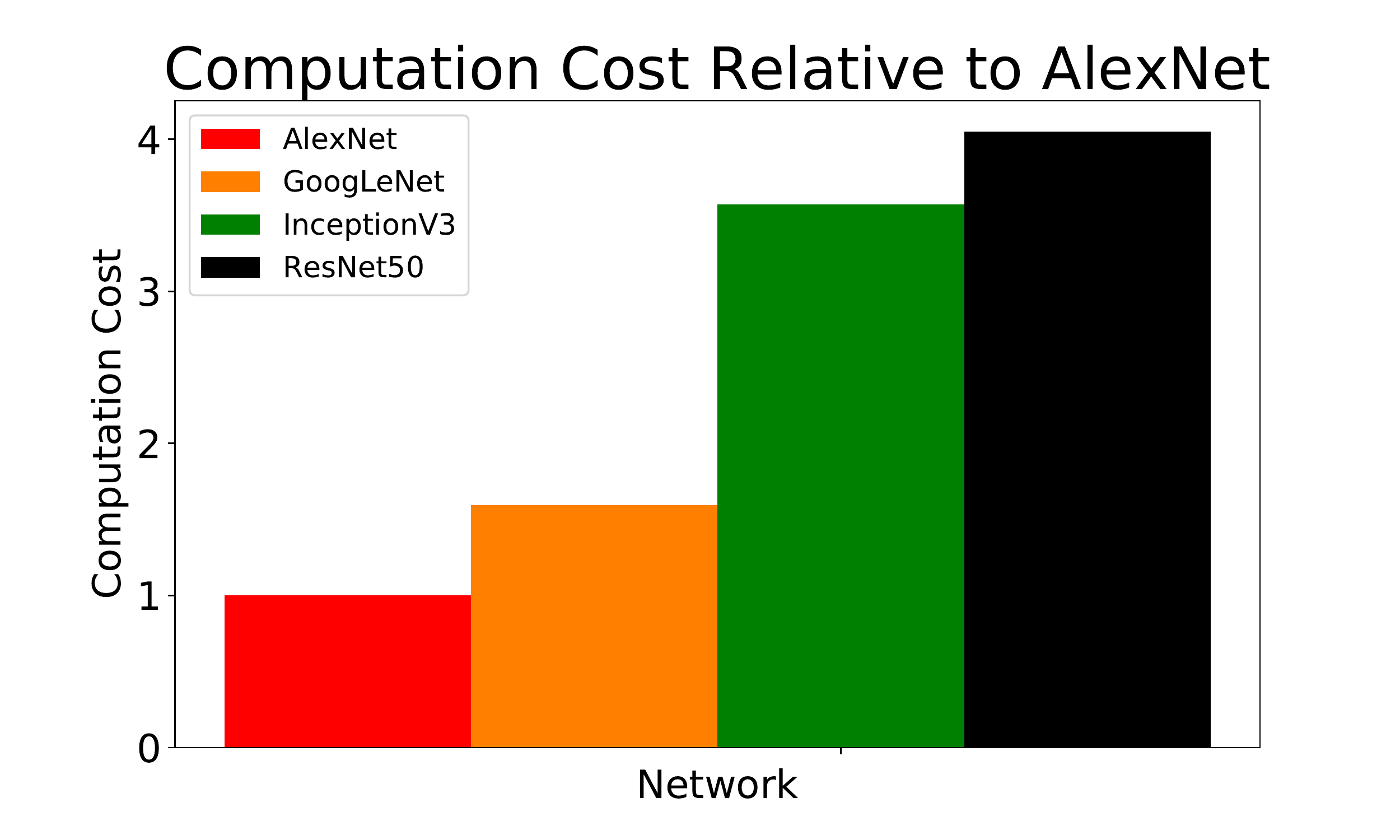}
\caption{Computation costs relative to AlexNet}
\label{fig:comp_costs}
\end{figure}

\begin{figure}[!htbp]
\includegraphics[width=\columnwidth]{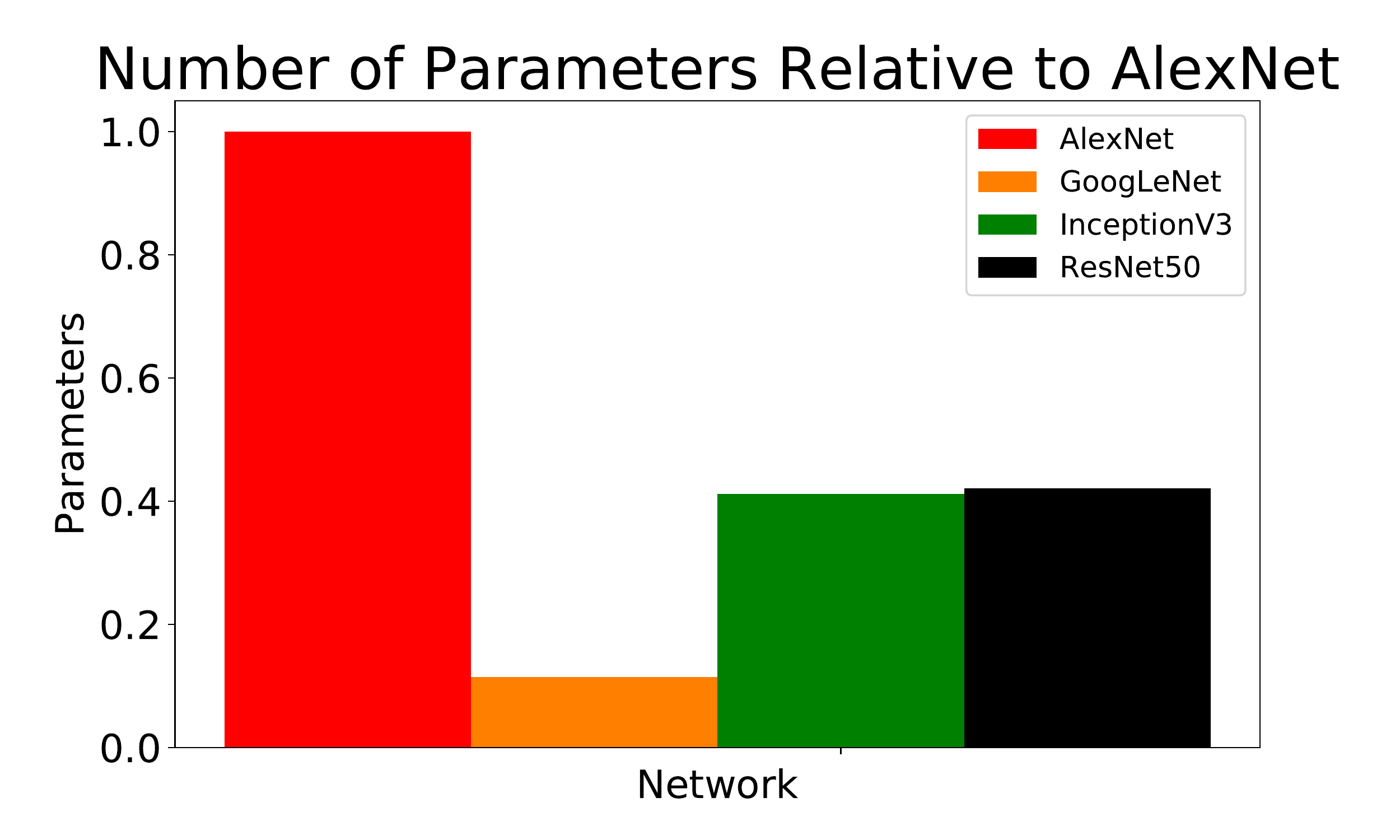}
\caption{Number of parameters relative to AlexNet}
\label{fig:comm_costs}
\end{figure}

\begin{figure}[!htbp]
\includegraphics[width=\columnwidth]{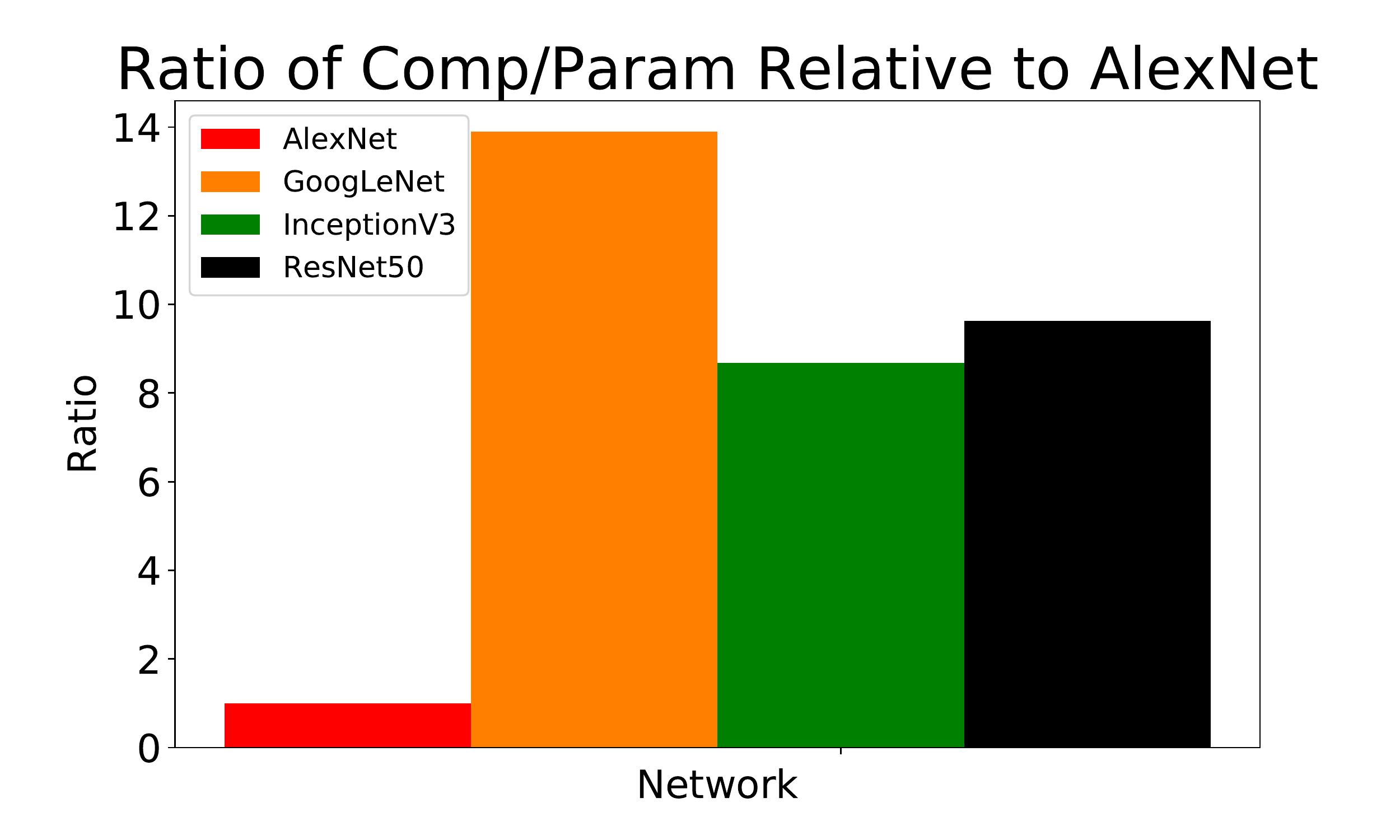}
\caption{Ratio of computation cost to parameters relative to AlexNet}
\label{fig:imagenet_ratio}
\end{figure}

In Figures~\ref{fig:comp_costs} and~\ref{fig:comm_costs} we evaluate both the
computation and communication costs of other neural networks relative to AlexNet -- the oldest of these
four models. These charts provide a graphical characterization of the scaling
potential for each network.  As the number of compute nodes increases, the
communication cost increases logarithmically, but the aggregate compute cost is
constant (under strong scaling).  This indicates that the models with a higher
ratio, as shown in Figure~\ref{fig:imagenet_ratio} will scale better. Based on these
figures, we see that the most difficult model to scale is AlexNet and the one
with the best scaling properties is GoogLeNet. This is empirically confirmed when examining
their performance with strong scaling experiments in section \ref{sec:eval}.


\subsection{Performance Comparisons}
\label{sec:eval}
\begin{figure}[!htbp]
\includegraphics[width=\columnwidth]{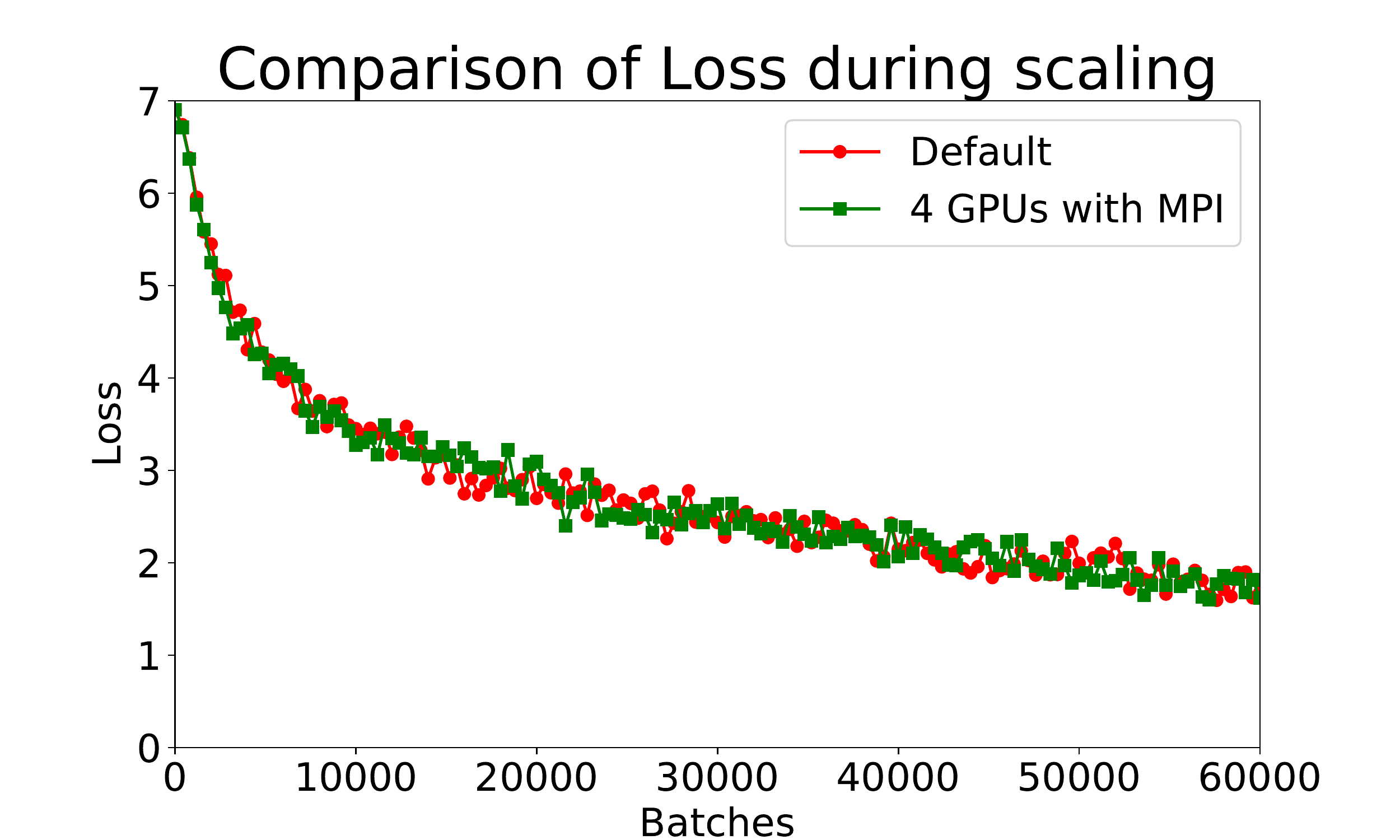}
\caption{Losses for default and 4 GPU training on AlexNet}
\label{fig:losses}
\end{figure}
\begin{figure*}[!htbp]
\subfloat[Speedup on \bf{SP}]{\includegraphics[width=\columnwidth]{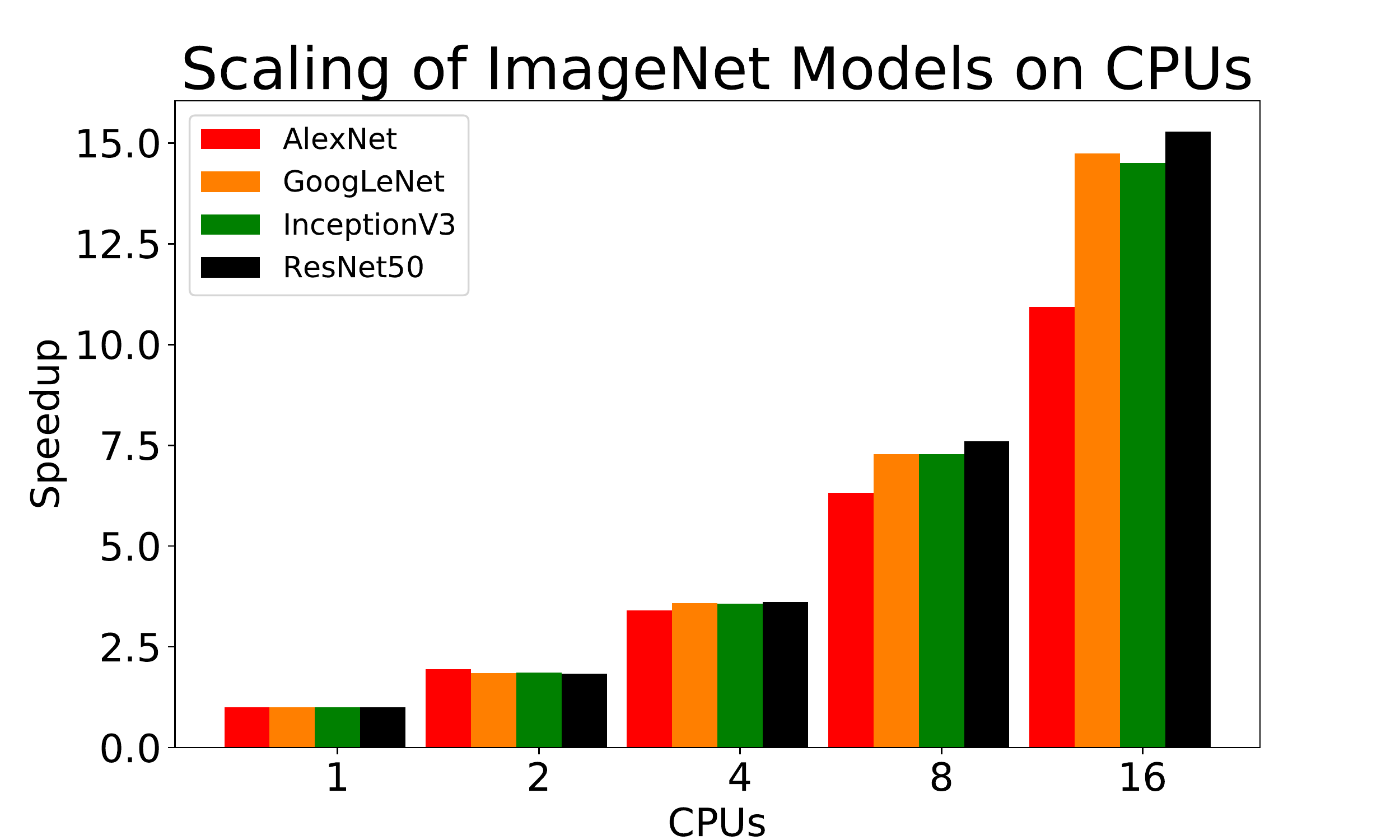}}
\subfloat[Speedup on \bf{K40}]{\includegraphics[width=\columnwidth]{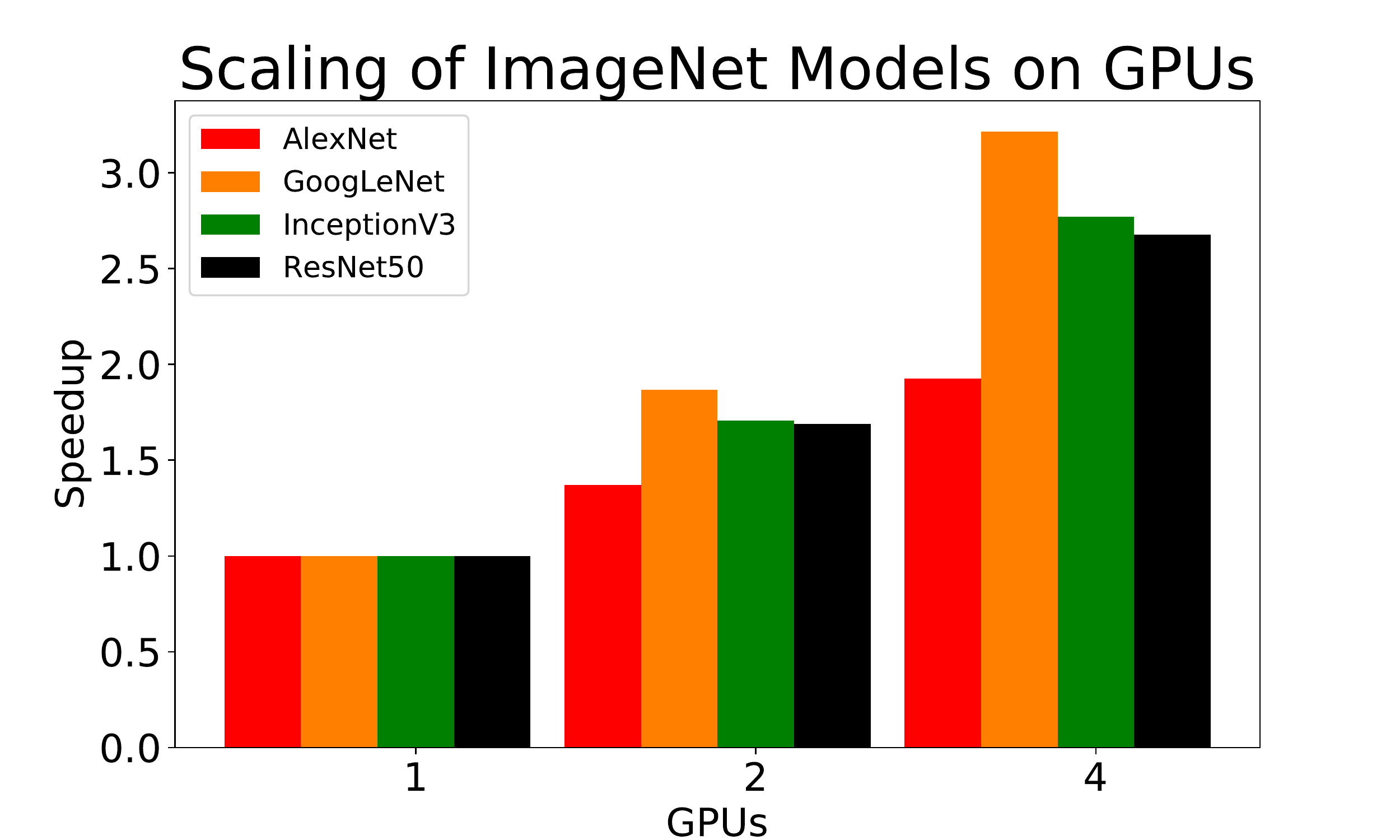}}
\caption{CPU and GPU speedups for four standard ImageNet classifiers relative to themselves on a single compute node}
\label{fig:imagenet_speedup}
\end{figure*}

In this section, we present a performance evaluation of MaTEx-TensorFlow using
several neural network models. We use SB and K40 architectures (please refer to
table~\ref{table:arch}).  Specifically, we present the speedup relative to 1
compute node/device (in the case of GPUs). We use strong scaling with a batch
size of 256 for AlexNet and GoogleNet, 128 for InceptionV3 and 64 for ResNet50.
Figure~\ref{fig:imagenet_speedup} shows the relative speedup comparisons for
CPU (SB architecture) and GPU (K40 architecture), respectively.  We observe
that AlexNet scales the worst of all achieving less than 2x speedup on 4 GPUs
and $\approx$ 11x speedup on 16 CPU nodes.  The ratio of computation to
communication dictates how well a network scales, with computationally more
expensive networks with fewer parameters, such as InceptionV3 and ResNet50
scaling better than AlexNet with GoogLeNet scaling the best on 4 K40 GPUs with
a speedup of $\approx 3.21x$). On CPUs, the tested (excluding AlexNet) models
scale well up to 16 CPU nodes, where GoogLeNet, InceptionV3 and ResNet50
respectively speedup by a factor of 14.7x, 14.5x and 15.3x, respectively.

We also note that with the addition of new user-operations, as described in
Section~\ref{sec:design} introduces non-trivial overhead.  We observe that the
overhead is $\approx$ 12\%. We intend to further reduce the overhead with upcoming releases of MaTEx-TensorFlow.

Figure~\ref{fig:losses} compares the loss curves of MaTEx-TensorFlow and
sequential TensorFlow using AlexNet. The objective is to empirically prove the
equivalence of MaTEx-TensorFlow in terms of loss in comparison to the
sequential implementation. We train AlexNet with a version of the quick solver
described in~\cite{siegel2016adaptive}. As observed from the figure, the losses are identical -- which validates our hypothesis.

\section{Related Work}
\label{sec:related}
Several researchers have conducted in-depth exploration of DL algorithms,
including a few focusing on multi-core/many-core systems.  Some of these
researchers further considered execution on large scale systems.  The most
widely used DL implementations include Caffe~\cite{jia2014caffe},
Warp-CTC~\cite{warpctc}, Theano~\cite{Bastien-Theano-2012,
bergstra+al:2010-scipy}, Torch~\cite{Collobert02torch:a}, Microsoft
CNTK~\cite{export:226641}, Chainer~\cite{chainer_learningsys2015} and Google
TensorFlow~\cite{tensorflow2015-whitepaper}, all of which implement GPU support
using NVIDIA CUDA Deep Neural Network (cuDNN) library.

For large scale execution of machine learning models in general, several
programming models have been proposed.  MapReduce~\cite{dean2008mapreduce}
provides large scale parallel execution using the Map and Reduce tasks.
Although MapReduce as a model is generic, its implementations, such as Hadoop,
have been widely critiques for performance reasons.  Spark, a recently proposed
programming model, supports in-memory iterative training of algorithms.
Distbelief~\cite{NIPS2012_0598} is an approach proposed by Dean {\em et al.},
using a parameter server for model updates at a central server, which despite
scaling well due to asynchronicity, has poor converge properties and the server
model becomes a bottleneck~\cite{Chen2016}.

Message Passing Interface (MPI)~\cite{mpi1,mpi2} has become the most common
method of building large scale DL algorithms.  It provides abstractions for
both pair-wise and group communication and is capable of using high speed
interconnects natively, making it particularly suitable to supercomputing
environments.  Among the toolkits that use MPI are Microsoft CNTK, the Machine
Learning Toolkit for Extreme Scale (MaTEx) version of
Caffe~\cite{siegel2016adaptive,vishnu:cluster15-a,vishnu:cluster15-b,vishnu:ipdps16,shohdy:icpp16,shohdy:hipc16,zheng:icpads16}, and the multi-node version of Chainer.

TensorFlow itself provides abstractions for building DL algorithms, including
computational graph structures and automatic differentiation.  Furthermore, it
provides methods for the user to define a parameter server style parallel
training regimen, using Google's Remote Procedure Call, which is restricted to
using sockets interface and static assignment of work to threads.  To do so,
the user must define a cluster, containing a server and workers, divide
communication tasks among them, specify that each device receives a copy of the
model, enforce synchronization, and wrap important operators so that the
parallel training can use them.  Similarly, a recent release by
Baidu~\cite{baiduflow}, which uses MPI to train a model in parallel, requires
that the user get MPI related variables from the environment, wrap the same
important operators as TensorFlow requires (along with several additional
ones).  Earlier work~\cite{1603.02339} included MPI outside of the TensorFlow
runtime, explicitly inserting the MPI commands into the user script.

\section{Acknowledgment}

This research and development is supported by a grant from Advanced Scientific
Computing Research (ASCR) on "Convergence of Machine Learning and Deep Learning
for HPC Modeling and Simulation", Analysis in Motion (AIM) Laboratory Directed
Research and Development (LDRD) and US Government.

\section{Conclusions}
\label{sec:conclusions}
Deep Learning (DL) algorithms have become a popular choice for data analysis. Several DL implementations -- primarily limited to
a single compute node -- such as Caffe, TensorFlow, Theano and Torch have become
readily available. Distributed DL implementations capable of execution on large
scale systems are becoming important to address the computational needs of
large data produced by scientific simulations and experiments.  Yet, the
adoption of distributed DL faces significant impediments: 1) Most
implementations require DL analysts to modify their code significantly -- which
is a show-stopper, 2) Several distributed DL implementations are
geared towards cloud computing systems -- which is inadequate for execution on
massively parallel systems such as supercomputers. 

This work addresses each of these problems. We provide a distributed memory DL
implementation by incorporating required changes in the TensorFlow runtime itself.
This dramatically reduces the entry barrier for using distributed TensorFlow
implementation.  We use Message Passing Interface (MPI) -- which provides
performance portability, especially since MPI specific changes are abstracted
from users. Lastly -- and arguably most importantly -- we make our
implementation available for broader use, under the umbrella of Machine
Learning Toolkit for Extreme Scale (MaTEx) at http://hpc.pnl.gov/matex.

\bibliographystyle{IEEEtran}
\balance
\bibliography{vishnu_papers,IEEEabrv,vishnu,vishnu2,vishnu_ml,FasterLearning,extra,psmo}
\end{document}